\documentclass[english,prb]{revtex4}
\usepackage[T1]{fontenc}
\usepackage[latin9]{inputenc}
\usepackage{amsmath}
\usepackage{graphicx}
\usepackage{amssymb}

\makeatletter
\def\vec#1{\mbox{\boldmath $#1$}}
\def\d{\textrm d}

\usepackage{babel}
\makeatother

\begin{document}

\title{Single-magnet rotary flowmeter for liquid metals}

\author{J\={a}nis Priede}

\affiliation{Applied Mathematics Research Centre, Coventry University, United
Kingdom}

\email{j.priede@coventry.ac.uk}

\author{Dominique Buchenau}

\author{Gunter Gerbeth}

\affiliation{Forschungszentrum Dresden-Rossendorf, MHD Department, Germany}

\begin{abstract}
We present a theory of single-magnet flowmeter for liquid metals and
compare it with experimental results. The flowmeter consists of a
freely rotating permanent magnet, which is magnetized perpendicularly
to the axle it is mounted on. When such a magnet is placed close to
a tube carrying liquid metal flow, it rotates so that the driving
torque due to the eddy currents induced by the flow is balanced by
the braking torque induced by the rotation itself. The equilibrium
rotation rate, which varies directly with the flow velocity and inversely
with the distance between the magnet and the layer, is affected neither
by the electrical conductivity of the metal nor by the magnet strength.
We obtain simple analytical solutions for the force and torque on
slowly moving and rotating magnets due to eddy currents in a layer
of infinite horizontal extent. The predicted equilibrium rotation
rates qualitatively agree with the magnet rotation rate measured on
a liquid sodium flow in stainless steel duct. 
\end{abstract}
\maketitle

\section{Introduction}

Flow rate measurements of liquid metals are required in various technological
processes ranging from the cooling of nuclear reactors to the dosing
and casting of molten metals.\cite{Cha-etal03} Electromagnetic flowmeters
are essential in the diagnostics and automatic control of such processes.
A variety of electromagnetic flowmeters have been developed starting
from the late 1940s and described by Shercliff.\cite{Sher62} The
standard approach is to determine the flow rate by measuring the potential
difference induced between a pair of electrodes by a flow of conducting
liquid in the magnetic field.\cite{Bev70,Hemp91} This approach is
now well developed and works reliably for common liquids like water,\cite{Fu-etal10}
but not so for liquid metals. Major problem in molten metals, especially
at elevated temperatures, is the electrode corrosion and other interfacial
effects, which can cause a spurious potential difference between the
electrodes.

The electrode problem is avoided by contactless eddy-current flowmeters,
which determine the flow rate by sensing the flow-induced perturbation
in an applied magnetic field.\cite{Feng75,SGG04} The main problem
with this type of flowmeters is the weak field perturbation which
may be caused not only by the flow. We showed recently that the flow-induced
phase shift of AC magnetic field is more reliable for flow rate measurements
than the amplitude perturbation.\cite{PBG10}

Another contactless techniques for flow rate measurements in liquid
metals is the so-called magnetic flywheel invented by Shercliff,\cite{Sher60}
who prescribes \emph{a {}``plurality''} of permanent magnets distributed
equidistantly along the circumference of a disk, which is mounted
on an axle and placed close to a tube carrying the liquid metal flow.
The eddy currents induced by the flow across the magnetic field interact
with the magnets by entraining them, which makes the disk rotate with
a rate proportional to that of the flow. This type of flowmeter, described
also in the textbook by Shercliff\cite{Sher62} and extensively used
by Bucenieks,\cite{Buc02,Buc05} was recently successfully reembodied
under the name of the Lorentz force velocimetry (LFV).\cite{TVK06}

Recently, we suggested an alternative and much more compact design
of such a flowmeter, which conversely to Shercliff's flywheel uses
just a single magnet mounted on the axle it can freely rotate around
and magnetized perpendicularly to it.\cite{GPBBGB-09} We also introduced
a basic mathematical model and presented first experimental implementation
of this type of flowmeter.\cite{PBG09} When such a magnet is placed
properly at a tube with the liquid metal flow, it starts to revolve
similarly to Shercliff's flywheel. But in contrast to the latter,
which is driven by the electromagnetic force acting on separate magnets,
the single magnet is set into rotation only by the torque. This driving
torque is due to the eddy currents induced by the flow across the
magnetic field. As the magnet starts to rotate, additional eddy currents
are induced, which brake the rotation. An equilibrium rotation rate
is attained when the braking torque balances the driving one, and
this rate depends only on the flow velocity and the flowmeter arrangement,
whereas it is independent of the electromagnetic torque itself. Thus,
the equilibrium rotation rate is affected neither by the magnet strength
nor by the electrical conductivity of the liquid metal provided that
the friction on the magnet is negligible. This a major advantage of
the single-magnet rotary flowmeter over the LFV approach, which relies
on direct force measurements.\cite{TVKZ-07}

In this paper, we present an extended theory of the single-magnet
rotary flowmeter and compare it with experimental results. Two limiting
cases of long and short magnets are analyzed using linear-dipole and
single-dipole approximations. We obtain simple analytic solutions
for the force and torque on slowly moving and rotating magnets due
to eddy currents in the layers of infinite horizontal extent and arbitrary
depth. This allows us to find the equilibrium rotation rate of the
magnet at which the torques due to the translation and rotation balance
each other. We also consider an active approach, where the force on
the magnet is used to control its rotation rate so that the resulting
force vanishes. This rotation rate, similarly to the equilibrium one,
is proportional to the layer velocity and independent of its conductivity
and the magnet strength.

The torque on a magnetic dipole rotating about an axis normal to a
thin sheet has been calculated by Smythe using an original receding
image method.\cite{Smythe-50} Reitz uses this method to calculate
the lift and drag forces on the coils of various geometries moving
with constant velocity above a conducting thin plate.\cite{Reitz-70}
The lift force, which at high speeds approaches the force between
the coil and its image located directly below it, varies as the velocity
squared in the low-speed limit considered in this paper. The drag
force is found to vary inversely and directly with the velocity at
low and high speeds, respectively. Palmer finds analytical expressions
for the eddy current forces on a circular current loop moving with
a constant velocity parallel to a thin conducting sheet.\cite{Palmer-04}
The force on a rectangular coil moving above a conducting slab has
been calculated numerically by Reitz and Davis using the Fourier transform
method.\cite{Reitz-Davis72} The same problem for the magnetic dipole
of arbitrary orientation placed next to a thin slowly moving slab
is addressed by Kirpo \emph{et al.} \cite{KBT10} The force and torque
on a transversely oriented dipole above a slowly moving plane layer
of arbitrary thickness has been found analytically in the context
of the LFV.\cite{TVKZ-07} Fast computation of forces on moving magnets
are of interested also for the eddy current force testing techniques.\cite{ZiBr-10}

This paper is organized as follows. The following section presents
two simple mathematical models of the single-magnet rotary flowmeter,
which are used to calculate analytically the force and torque on the
magnets moving and rotating slowly above a layer of infinite lateral
extent. The limits of long and short magnets, which are approximated
by linear and point dipoles, are considered in Secs. \ref{sub:2D}
and \ref{sub:3D}, respectively. Section \ref{sec:Exp} presents the
flowmeter implementation details and test results. The paper is concluded
by a summary in Sec. \ref{sec:Con}.

\section{Theory}

\subsection{Formulation of problem}

\begin{figure}
\begin{centering}
\includegraphics[width=0.33\columnwidth]{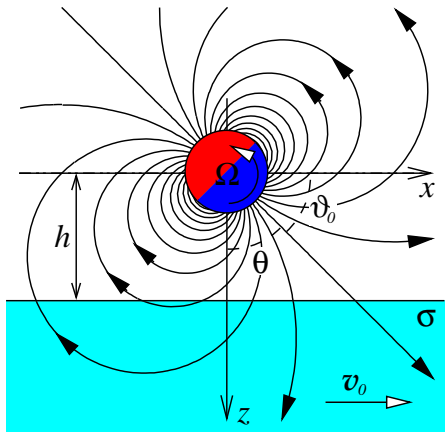} 
\par\end{centering}

\caption{\label{cap:Sketch}Schematic view of the single-magnet rotary flowmeter
with the $y$-axis directed out of the plane of figure. }

\end{figure}

Consider a horizontally unbounded planar layer of electrical conductivity
$\sigma$ occupying the lower half-space and moving as a solid body
with a constant velocity $\vec{v}_{0}$ parallel to a permanent magnet
placed at a distance $h$ above its surface and rotating with a constant
angular velocity $\vec{\Omega}$ around an axis parallel to the surface
and perpendicular to $\vec{v}_{0}.$ Velocities are assumed sufficiently
low for the magnetic field of induced currents to be negligible compared
to the field of the magnet. The origin of Cartesian coordinates is
set at the center of the magnet with the $x,$ $y$ and $z$ axis
directed along $\vec{v}_{0}=v\vec{e}_{x},$ $\vec{\Omega}=\Omega\vec{e}_{y},$
and downward normally to the surface, respectively, as shown in Fig.
\ref{cap:Sketch}. In the following, two limiting cases will be considered
in which the magnet will be assumed either much longer or much shorter
than $h.$

\subsection{\label{sub:2D}Linear-dipole model for a long magnet}

We start with a long cylinder magnetized perpendicularly to its axis
about which it can freely rotate.\cite{PBG09} In this case, the
magnetic field is approximated by that of a two-dimensional (linear)
dipole with the vector potential $\vec{A}\left(\vec{r}\right)=\vec{e}_{y}A\left(\vec{r}\right),$
which has only the $y$-component \begin{equation}
A\left(\vec{r}\right)=\frac{\mu_{0}}{2\pi}\frac{\vec{\bar{m}}\cdot\vec{r}}{r^{2}}=\frac{\mu_{0}}{2\pi}\frac{\bar{m}\cos(\vartheta-\vartheta_{0})}{r},\label{eq:A2d}\end{equation}
 where $\mu_{0}=4\pi\times10^{-7}\, H/m$ is the vacuum permeability
and $\vec{\bar{m}}$ is the linear dipole moment, which is perpendicular
to $\vec{e}_{y}$ and directed at the angle $\vartheta_{0}$ from
the positive $x$ axis; $\vec{r}$ is the radius vector from the magnet
axis, and $r=\left|\vec{r}\right|$ and $\vartheta$ are the cylindrical
radius and the polar angle in the cylindrical coordinates around the
$y$-axis. The magnetic field of linear dipole is \begin{equation}
\vec{B}=\vec{\nabla}\times\vec{A}=-\vec{e}_{y}\times\vec{\nabla}A.\label{eq:B2d}\end{equation}
 Eddy currents are induced by two effects: the translation of the
layer and the temporal variation of the magnetic field due to its
rotation. The latter vanishes in the co-rotating frame of reference,
where the magnetic field is stationary, while the layer appears to
move with the resulting velocity $\vec{v}=\vec{v}_{0}+\vec{v}_{1},$
which contains also an apparent rotational motion of the layer $\vec{v}_{1}=-\vec{\Omega}\times\vec{r}$
opposite to that of the magnet. The density of eddy currents is given
by Ohm's law for a moving medium \begin{equation}
\vec{j}=\sigma(-\vec{\nabla}\varphi+\vec{v}\times\vec{B})\label{eq:Ohm}\end{equation}
 where $\varphi$ is the electric potential. In this case, no electric
potential is induced because the e.m.f., $\vec{v}\times\vec{B,}$
is both solenoidal and tangential to the surface. If the induced magnetic
field is negligible as originally assumed, eddy currents can be represented
as the superposition \[
\vec{j}=(j_{0}+j_{1})\vec{e}_{y},\]
 where $j_{0}=-\sigma\vec{v}_{0}\cdot\vec{\nabla}A$ and $j_{1}=-\sigma\vec{v}_{1}\cdot\vec{\nabla}A$
are the currents induced by the translation and rotation, respectively.

It is important to note that in the approximation under consideration
with a fixed magnetic field distribution, eddy currents are determined
only by the instantaneous velocities. Besides that eddy currents are
coplanar to the surface and, thus, mutually independent over the depth
of the layer. Consequently, a layer of finite thickness may be represented
as a semi-infinite one with zero velocity at $z>h_{2},$ where $h_{2}$
is the distance of the lower boundary of finite-thickness layer from
the magnet. This, in turn, is equivalent to the superposition of two
semi-infinite layers with the second layer at $z>h_{2}$ moving oppositely
to the first one at $z>h_{1}$ so that the resulting velocity vanishes
at $z>h_{2}.$ In the following, this approach allows us to construct
the solution for a finite-thickness layer by taking the difference
of two half-space solutions, which is subsequently denoted by $\left[X\right]_{h_{1}}^{h_{2}},$
where $X$ stands either for the force or the torque due to the eddy
currents in a half-space. Moreover, by the same arguments, this approach
can easily be extended to $z$-dependent velocity distributions $v(z)$,
for which general solution can be constructed as a superposition of
solutions for thin layers moving with various velocities given by
$\int v(z)\partial_{z}X\,\d z,$ where $\partial_{z}X=\left[X\right]_{h_{1}}^{h_{2}}/\left[z\right]_{h_{1}}^{h_{2}}$
for $h_{2}\rightarrow h_{1}=z$ and $v=1.$

The linear force density experienced by an infinitely long magnet
due to the layer translation, which according to the momentum conservation
law is opposite to that acting upon the layer itself, can be written
as $-\int_{S}\vec{j}_{0}\times\vec{B}\,\d s=F_{0}\vec{e}_{x},$ where
the integral is taken over the $xz$-cross-section of the layer. The
$y$-component of force is absent due to the $y\rightarrow-y$ reflection
symmetry. In the low-speed limit under consideration, when force varies
linearly with the velocity, there is also no $z$-component of force.
This is the case because according to the linearity $\vec{F}\rightarrow-\vec{F}$
when $\vec{v}_{0}\rightarrow-\vec{v}_{0},$ while the latter transformation
is equivalent to the rotation of the coordinate system by $180^{\circ}$
around the $z$-axis, which leaves the $z$-components invariant.
In polar coordinates with the surface defined by $r_{1}=h/\sin\vartheta,$
we obtain \begin{equation}
F_{0}=-\frac{\mu_{0}^{2}\bar{m}^{2}\sigma v}{4\pi^{2}}\int_{\pi}^{2\pi}\cos^{2}(2\vartheta-\vartheta_{0})\int_{r_{1}}^{\infty}\frac{\d r}{r^{3}}\,\d\vartheta=\frac{\mu_{0}^{2}\bar{m}^{2}\sigma v}{32\pi h^{2}}.\label{eq:F0-gen}\end{equation}
 The linear torque density, which because of the aforementioned symmetries
has only the $y$-component, can be found as $-\int_{S}\vec{r}\times\vec{j}_{0}\times\vec{B}\,\d s=M_{0}\vec{e}_{y},$
where \begin{equation}
M_{0}=\frac{\mu_{0}^{2}\bar{m}^{2}\sigma v}{4\pi^{2}}\int_{\pi}^{2\pi}\cos(2\vartheta-\vartheta_{0})\sin(\vartheta-\vartheta_{0})\int_{r_{1}}^{\infty}\frac{\d r}{r^{2}}\,\d\vartheta=\frac{\mu_{0}^{2}\bar{m}^{2}\sigma v}{16\pi h}.\label{eq:M01-gen}\end{equation}
 The linear force density due to the rotation, defined by $-\int_{S}\vec{j}_{1}\times\vec{B}\,\d s=F_{1}\vec{e}_{x},$
is found as \[
F_{1}=\sigma\int_{S}(\vec{e}_{x}\cdot\vec{\nabla}A)(\vec{v}_{1}\cdot\vec{\nabla}A)\,\d s=-\frac{\mu_{0}^{2}\bar{m}^{2}\sigma\Omega}{16\pi h}.\]
 The linear torque density due to the rotation, as that due to the
translation above, has only the $y$-component \begin{equation}
M_{1}=\sigma\int_{S}(\vec{r}\times\vec{\nabla}A)(\vec{v}_{1}\cdot\vec{\nabla}A)\cdot\d\vec{s}=\frac{\mu_{0}^{2}\bar{m}^{2}\sigma\Omega}{4\pi^{2}}\int_{0}^{\pi}\sin^{2}(\vartheta-\vartheta_{0})\,\d\vartheta\int_{r_{1}}^{\infty}\frac{\d r}{r},\label{eq:M1-gen}\end{equation}
 which is not defined for a semi-infinite layer because the last integral
diverges. Nevertheless, expression (\ref{eq:M1-gen}) can still be
evaluated for the layer of finite depth by substituting the infinite
limit in the last integral by $r_{2}=h_{2}/\sin\vartheta,$ which
results in \begin{equation}
\frac{\mu_{0}^{2}\bar{m}^{2}\sigma\Omega}{8\pi}\ln\frac{h_{2}}{h_{1}}=\left[M_{1}\right]_{h_{2}}^{h_{1}}.\label{eq:M1-h2}\end{equation}

The solution above becomes unbounded as $h_{2}\rightarrow\infty$
and, thus, inapplicable to thick layers. This implies that for a half-space
layer the induced magnetic field cannot be neglected however slow
the rotation of magnet. The induced magnetic field is related to eddy
currents by Ampere's law $\vec{j}=\frac{1}{\mu_{0}}\vec{\nabla}\times\vec{B},$
which combined with expressions (\ref{eq:B2d}) and (\ref{eq:Ohm})
leads to \begin{equation}
\mu_{0}\sigma\vec{v}_{1}\cdot\vec{\nabla}A+\vec{\nabla}^{2}A=0.\label{eq:A}\end{equation}
 As suggested by the external magnetic field (\ref{eq:A2d}), we search
for the vector potential in the complex form \begin{equation}
A(\vec{r})=\Re\left[\hat{A}(r)e^{i(\vartheta-\vartheta_{0})}\right].\label{eq:A-cmp}\end{equation}
 Then Eq. (\ref{eq:A}) for the complex amplitude $\hat{A}$ takes
the form \begin{equation}
-i\mu_{0}\sigma\Omega\hat{A}=r^{-1}(r\hat{A}')'-r^{-2}\hat{A},\label{eq:A-rot}\end{equation}
 where the prime stands for the derivative with respect $r.$ The
general solution of Eq. (\ref{eq:A-rot}) can be written as $\hat{A}(r)=CK_{1}\left((1+i)r/d\right),$
where $C$ is an unknown constant to be determined by matching the
induced and externally imposed magnetic fields; $K_{1}(z)$ is the
modified Bessel function of the second kind,\cite{AbSt72} and $d=\sqrt{2/(\mu_{0}\sigma\Omega)}$
is the skin depth due to the rotation. At distances $r$ smaller than
the skin depth $(r/d\ll1),$ $\hat{A}\approx Cd/((1+i)r)$ is expected
to approach the imposed field $\hat{A}=\mu_{0}\bar{m}/(2\pi r),$
which yields $C=(1+i)\mu_{0}\bar{m}/(2\pi d$) and \begin{equation}
\hat{A}(r)=\frac{(1+i)\mu_{0}\bar{m}}{2\pi d}K_{1}\left((1+i)r/d\right).\label{eq:A-hat}\end{equation}
 Substituting expressions (\ref{eq:A-cmp}) and (\ref{eq:A-hat})
into integral (\ref{eq:M1-gen}), the torque on the magnet can be
represented as $M_{1}=\frac{\sigma\Omega}{2}\int_{0}^{\pi}I(r_{1}(\vartheta))\,\d\vartheta,$
where $I(r)=\int_{r}^{\infty}\Re\left[\hat{A}^{*}\hat{A}\right]r\,\d r$
and the asterisk denotes the complex conjugate. Using Eq. (\ref{eq:A-rot}),
after some algebra we obtain \[
I(r)=\frac{d^{2}}{2}r\Im\left[\hat{A}^{*}\hat{A}'\right]=\left(\frac{\mu_{0}}{2\pi}\right)^{2}\Re\left[K_{0}((1+i)r/d)K_{1}((1-i)r/d)(1+i)r/d\right].\]
 For the low rotation rates that satisfy $r/\delta\ll1$, we obtain
\[
I(r)\approx-\left(\frac{\mu_{0}}{2\pi}\right)^{2}\Re\left[\ln((1-i)r/d)+\gamma\right]=\left(\frac{\mu_{0}}{2\pi}\right)^{2}(\ln(2d/r)-\gamma),\]
 where $\gamma=0.577215\ldots$ is Euler's constant. Finally, taking
into account that $r_{1}(\vartheta)=h_{1}/\sin\vartheta$ and $\int_{0}^{\pi}\ln(\sin\vartheta)\,\d\vartheta=-\pi\ln2,$
we obtain\begin{equation}
M_{1}=\frac{\mu_{0}^{2}\bar{m}^{2}\sigma\Omega}{8\pi}\ln\frac{\tilde{d}}{h_{1}},\label{eq:M1-skin}\end{equation}
 where $\tilde{d}=de^{-\gamma}$ is the effective skin depth, which
owing to the low velocities under consideration is supposed to be
large relative to $h_{1}$.

Now we can use the results above to find the magnet rotation rate
depending on the velocity of layer. First, if the magnet rotates steadily
without a significant friction, the torques due to the translation
and rotation are at equilibrium: $\left[M_{0}+M_{1}\right]_{h_{2}}^{h_{1}}=0,$
which yields \begin{equation}
\Omega=\frac{v}{2}\left(\frac{1}{h_{1}}-\frac{1}{h_{2}}\right)/\ln\frac{h_{2}}{h_{1}}.\label{eq:omg-trq}\end{equation}
 Note that this equilibrium rotation rate $\Omega$ depends neither
on the magnet strength nor on the layer conductivity unless $h_{2}\gtrsim\tilde{d}.$
In the latter case, the skin effect becomes important and, thus, $h_{2}$
has to be substituted by $\tilde{d}$ in the expression above. 

The velocity of layer can be determined also in another way by measuring
the force on the magnet, as in the LFV, to control the magnet rotation
rate so that the resulting force vanishes, $\left[F_{0}+F_{1}\right]_{h_{2}}^{h_{1}}=0.$
This results in\begin{equation}
\Omega=\frac{v}{2}\left(\frac{1}{h_{1}}+\frac{1}{h_{2}}\right),\label{eq:omg-frc}\end{equation}
 which again depends linearly on the layer velocity, but does not
depend on its conductivity or the magnet strength.

\subsection{\label{sub:3D}Single-dipole model for a short magnet}

In the other limiting case of a short magnet, when the distance to
the surface is large or at least comparable to the size of magnet,
the latter can be considered as a dipole with the scalar magnetic
potential \begin{equation}
\Phi(\vec{x};\vec{m})=-\vec{m}\cdot\vec{\nabla}G(\vec{x}),\label{eq:Phi}\end{equation}
 where $\vec{m}$ is the dipole moment and $G(\vec{x})=(4\pi|\vec{x}|)^{-1}$
is the fundamental solution of Laplace's equation, which satisfies
$\vec{\nabla}^{2}G=-\delta(\vec{x})$ with the Dirac delta function
on the r.h.s. In the following, we use simplified notation $\vec{e}_{m}\cdot\vec{\nabla}\equiv\partial_{m},$
where $\vec{e}_{m}=\vec{m}/m$ is the unit vector. Then the dipole
magnetic field is given by $\vec{B}=-\mu_{0}(\vec{\nabla}\Phi-\delta(\vec{x})\vec{m}),$
where the last term is added to ensure the solenoidality of $\vec{B}$
also at $\vec{x}=0.$ The associated dipole current distribution is
\begin{equation}
\vec{J}=\frac{1}{\mu_{0}}\vec{\nabla}\times\vec{B}=\vec{\nabla}\times\delta\vec{m},\label{eq:J}\end{equation}
 which easily leads to the classical expressions for the force and
torque used later on. In the following, we will be using also the
spherical and cylindrical coordinates associated with the Cartesian
ones in the usual way. As in the previous section, we change to the
co-rotating frame of reference and consider the eddy currents due
to translation and rotation separately.

\subsubsection{Translation}

For the translation with $\vec{v}_{0}=v\vec{e}_{x},$ the charge conservation
$\vec{\nabla}\cdot\vec{j}=0$ applied to Eq. (\ref{eq:Ohm}) results
in Laplace's equation for $\varphi_{0}$\begin{equation}
\vec{\nabla}^{2}\varphi_{0}=0.\label{eq:phi0}\end{equation}
 At the surface $z=h,$ the normal component of electric current vanishes:
\begin{equation}
\partial_{z}\varphi_{0}=vB_{y}.\label{bc:phi0}\end{equation}
 In order to find the induced electric potential, firstly, it is important
to notice that $\vec{B}$ being a free-space magnetic field satisfies
the Laplace equation itself. Consequently, the Cartesian components
of $\vec{B}$ satisfy this equation, too. Secondly, if $B_{y}$ in
BC (\ref{bc:phi0}) satisfies Laplace's equation, then \begin{equation}
\varphi_{0}=v\int B_{y}\,\d z=\mu_{0}mv\int\partial_{ym}^{2}G\,\d z.\label{sol:phi0-gen}\end{equation}
 satisfies not only BC (\ref{bc:phi0}) but also Eq. (\ref{eq:phi0})
because the integration along a straight line, similarly to the differentiation,
are interchangeable with the Laplacian. By the same argument, we can
interchange integration and differentiation in expression (\ref{sol:phi0-gen}),
which yields \begin{equation}
\varphi_{0}=\mu_{0}mv\partial_{ym}^{2}H,\label{sol:phi0-H}\end{equation}
where $H=\int G\,\d z=\frac{1}{4\pi}(Q_{0}(\cos\theta)+q(r))$ and
$Q_{0}(z)=\ln\sqrt{\frac{1+z}{1-z}}$ is the zeroth degree associated
Legendre function of the second kind;\cite{AbSt72} $\theta$ is
the spherical polar angle from the positive $z$-axis (see Fig. \ref{cap:Sketch})
and $r$ is the corresponding cylindrical radius. The ``constant''
of integration $q(r),$ which similarly to the first term $Q_{0}(\cos\theta)$
is supposed to be axisymmetric and also to satisfy the Laplace equation,
is chosen to regularize $H$ at $r\rightarrow0$ by removing the logarithmic
singularity $Q_{0}=\ln\cot\frac{\theta}{2}\sim-\ln r$ for $z>0.$
This results in $q(r)=\ln r$ and\begin{equation}
H=\frac{1}{4\pi}\left(Q_{0}(z/R)+\ln r\right),\label{eq:H}\end{equation}
 where $R=|\vec{x}|$ is the spherical radius. For a transversal dipole
$(\vec{e}_{m}=\vec{e}_{z}),$ considered also by Thess \emph{at al.}\cite{TVKZ-07},
solution (\ref{sol:phi0-H}) reduces to \begin{equation}
\varphi_{0}=\mu_{0}vm\partial_{y}G=-\mu_{0}v\Phi(\vec{x};m\vec{e}_{y}),\label{sol:phi0-z}\end{equation}
 where the last term represents the magnetic potential of a dipole
aligned with the $y$-axis. For a general dipole orientation, expression
(\ref{eq:H}) substituted into solution (\ref{sol:phi0-H}) after
some algebra yields\begin{equation}
\varphi_{0}=\frac{\mu_{0}v}{4\pi}\left[\left(\frac{R-z}{Rr^{2}}-\frac{z}{2R^{3}}\right)(m_{x}\sin2\vartheta-m_{y}\cos2\vartheta)+\frac{2m_{z}r\sin\vartheta-m_{y}z}{2R^{3}}\right],\label{sol:phi0}\end{equation}
 where $\vartheta$ is the azimuthal angle from the positive $x$-axis
in the $xy$-plane.

The electric potential distribution (\ref{sol:phi0}) allows us to
calculate the force acting upon the magnet, which, as noted above,
is opposite to that acting upon the layer, i.e., $\vec{F}=-\int_{V}\vec{f}\,\d V,$
where the integral of the Lorentz force density $\vec{f}=\vec{j}\times\vec{B}$
is taken over the layer volume $V.$ For the longitudinal force component,
we obtain\begin{eqnarray}
F_{0,x} & = & \mu_{0}^{2}m^{2}\sigma v\int_{S}\left\{ \frac{1}{2}\partial_{z}\left[(\partial_{ym}^{2}H)^{2}-(\partial_{m}G)^{2}\right]+\int_{h}^{\infty}(\partial_{xm}^{2}G)^{2}\,\d z\right\} \,\d s\nonumber \\
 & = & \frac{\mu_{0}^{2}\sigma v}{512\pi h^{3}}(3m_{x}^{2}+m_{y}^{2}+4m_{z}^{2}),\label{eq:Fx}\end{eqnarray}
 where the first integral is taken over the surface $S$ at $z=h$
and can be swapped with the second one over the layer depth. For a
transversal dipole $(m_{x}=m_{y}=0)$, the expression above coincides
with that of Thess \emph{et al.}\cite{TVKZ-07} as well as with the
result of Reitz\cite{Reitz-70} for a thin sheet in the limit of
a slowly moving dipole. As seen, the force is the strongest on a transversal
dipole and reduces on longitudinal and spanwise dipoles by factors
of $\frac{3}{4}$ and $\frac{1}{4}$, respectively. Thus, the force
(\ref{eq:Fx}), in contrast to force (\ref{eq:F0-gen}) for a long
magnet, varies with the dipole orientation in the $xz$-plane. On
a horizontally inclined dipole $(m_{x}m_{y}\not=0)$, there is also
a spanwise force component\begin{eqnarray}
F_{0,y} & = & \mu_{0}^{2}m^{2}\sigma v\int_{S}\left\{ (\partial_{ym}^{2}H)(\partial_{xy}^{2}G)+\int_{h}^{\infty}(\partial_{xm}^{2}G)(\partial_{ym}^{2}G)\,\d z\right\} \,\d s\nonumber \\
 & = & \frac{\mu_{0}^{2}\sigma v}{256\pi h^{3}}m_{x}m_{y}.\label{eq:Fy}\end{eqnarray}
 But there is no vertical force whatever the dipole orientation:\begin{equation}
F_{0,z}=\mu_{0}^{2}\sigma v\int_{h}^{\infty}\int_{S}(\partial_{xm}^{2}G)(\partial_{zm}^{2}G)\,\d s\,\d z=0.\label{eq:Fz}\end{equation}
 This is consistent with the result of Reitz\cite{Reitz-70} stating
that in the low-speed limit the lift force is proportional to the
velocity squared, while the approximation under consideration takes
into account only the part of force proportional to the velocity. 

Alternatively, the force and torque acting on dipole can be found
using the associated current distribution (\ref{eq:J}) and the induced
magnetic field $\vec{b},$ which lead straightforwardly to the classical
expressions\cite{Jac98} \begin{eqnarray}
\vec{F} & = & \int_{\bar{V}}\vec{J}\times\vec{b}\,\d^{3}\vec{x}=(\vec{m}\cdot\vec{\nabla})\vec{b},\label{eq:F}\\
\vec{M} & = & \int_{\bar{V}}\vec{x}\times\vec{J}\times\vec{b}\,\d^{3}\vec{x}=\vec{m}\times\vec{b},\label{eq:M}\end{eqnarray}
 where the integrals are taken over the space $\bar{V}$ above the
layer. This is a bit longer but algebraically more straightforward
approach, which will be pursued in the following.

In order to find the induced magnetic field $\vec{b},$ it is important
to notice that solution (\ref{sol:phi0-gen}) satisfies condition
(\ref{bc:phi0}) not only at $z=h$ but at any $z.$ Thus, the $z$-component
of the current is absent not only at the surface but throughout the
whole layer. Then $\vec{\nabla}\cdot\vec{j}=0$ and $j_{z}\equiv0$
imply \begin{equation}
\vec{j}=\vec{\nabla}\times\psi\vec{e}_{z},\label{eq:j-psi}\end{equation}
 where $\psi$ is the electric stream function, whose isolines coincide
with the eddy current lines. Substituting this expression into Eq.
(\ref{eq:Ohm}) and taking the $z$-component of the \emph{curl} of
the resulting equation, we obtain\begin{equation}
(\vec{\nabla}^{2}-\partial_{z}^{2})\psi=-\sigma\vec{e}_{z}\cdot\vec{\nabla}\times(\vec{v}\times\vec{B}).\label{eq:psi-gen}\end{equation}
 Note that in contrast to the electric potential in Eq. (\ref{eq:phi0}),
no boundary conditions are required for $\psi$ at the surface. This
is because, firstly, Eq. (\ref{eq:psi-gen}) contains no derivatives
in $z$ and, secondly, the absence of the $z$-component of current
at the surface is explicitly ensured by expression (\ref{eq:j-psi}).

\begin{figure}
\begin{centering}
\includegraphics[width=0.5\columnwidth]{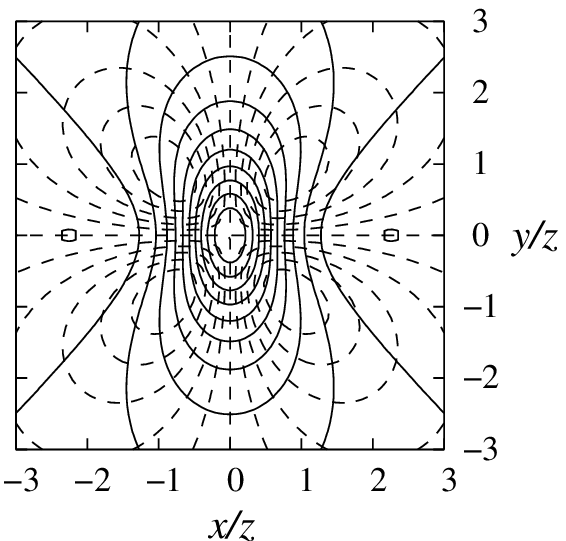}\put(-30,0){(a)}\includegraphics[width=0.5\columnwidth]{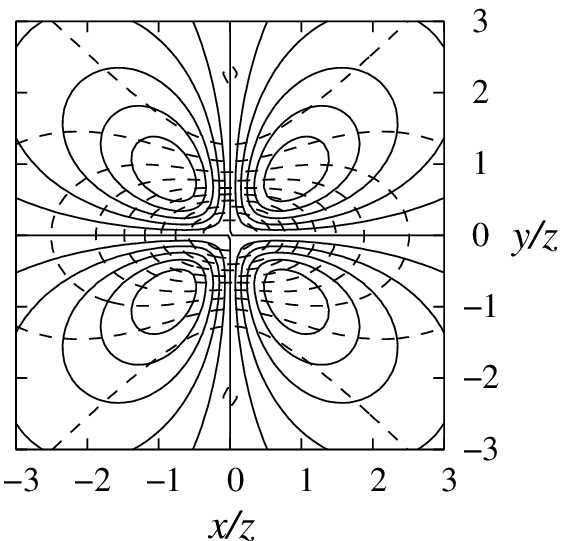}\put(-30,0){(b)}
\par\end{centering}

\begin{centering}
\includegraphics[width=0.5\columnwidth]{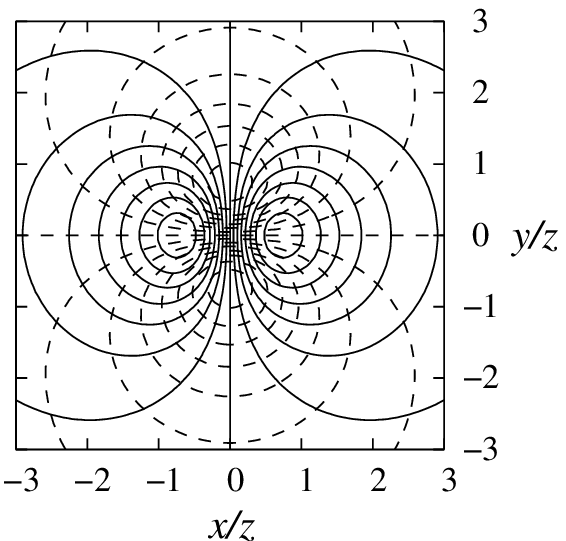}\put(-30,0){(c)} 
\par\end{centering}

\caption{\label{fig:mov-xyz}Isolines of the electric potential $\varphi_{0}$
(dashed) and of the stream function $\psi_{0}$ (solid) in the $xy$-plane,
where the latter represent the eddy current lines, induced by translation
in the magnetic field of the dipole aligned with the $x$- (a), $y$-
(b), and $z$-axis (c). }

\end{figure}

For translational motion with $\vec{v}_{0}=v\vec{e}_{x},$ the r.h.s
of Eq. (\ref{eq:psi-gen}) takes the form $\sigma v\partial_{x}B_{z}=-\mu_{0}\sigma v\partial_{xz}^{2}\Phi,$
which by the same arguments as above satisfies Laplace's equation
and equals to $-\partial_{z}^{2}\psi_{0}$ with \begin{equation}
\psi_{0}=-\mu_{0}m\sigma v\partial_{xm}^{2}H,\label{eq:psi0-H}\end{equation}
which satisfies the same equation and, thus, represents the solution
of Eq. (\ref{eq:psi-gen}). For a transversal dipole $(\vec{e}_{m}=\vec{e}_{z}$),
the general solution (\ref{eq:psi0-H}) simplifies to \begin{equation}
\psi_{0}=-\mu_{0}m\sigma v\partial_{x}G=\mu_{0}\sigma v\Phi(\vec{x};m\vec{e}_{x}),\label{sol:psi0-z}\end{equation}
 where the last term represents the magnetic potential of a longitudinal
dipole. Thus, in this case, the eddy current lines coincide with the
isolines of electric potential (\ref{sol:phi0}) rotated by $90^{\circ}$
about the $z$-axis. For a longitudinal dipole $(\vec{e}_{m}=\vec{e}_{x}),$
the comparison of solutions (\ref{eq:psi0-H}) and (\ref{sol:phi0-H})
shows that the eddy current lines and the electric potential isolines
are swapped with the corresponding distributions induced by a spanwise
dipole $(\vec{e}_{m}=\vec{e}_{y})$ and rotated by $90^{\circ}$ about
the $z$-axis. For an arbitrarily oriented dipole, expression (\ref{eq:H})
substituted into the general solution (\ref{eq:psi0-H}) after some
algebra yields\begin{equation}
\psi_{0}=-\frac{\mu_{0}\sigma v}{4\pi}\left[\left(\frac{R-z}{Rr^{2}}-\frac{z}{2R^{3}}\right)(m_{x}\cos2\vartheta+m_{y}\sin2\vartheta)+\frac{2m_{z}r\cos\vartheta-m_{x}z}{2R^{3}}\right].\label{sol:psi0}\end{equation}
Figure \ref{fig:mov-xyz} shows the isolines of the electric potential
(\ref{sol:phi0}) and those of $\psi_{0},$ which represent the eddy
current lines, in the $xy$-plane $(z=\mbox{const})$ for three basic
dipole orientations along the $x$-, $y$- and $z$-axis. Note that
the patterns are self-similar in the $xy$-plane with the characteristic
length scaling directly with $z.$ Namely, these distributions are
functions of spatial angles only, while according to (\ref{sol:phi0})
and (\ref{sol:psi0}) the magnitude of both $\varphi_{0}$ and $\psi_{0}$
falls off as $\sim z^{-2}.$

In order to satisfy the solenoidality condition $\vec{\nabla}\cdot\vec{b}=0,$
the induced magnetic field is sought as $\vec{b}=\vec{\nabla}\times\vec{a},$
where $\vec{a}$ is the vector potential, which is supposed to satisfy
the Coulomb gauge $\vec{\nabla}\cdot\vec{a}=0.$ Then Ampere's law
leads to $\vec{\nabla}^{2}\vec{a}=-\mu_{0}\vec{j},$ which, in turn,
results in $\vec{a}(\vec{x})=\mu_{0}\int_{V}\vec{j}(\vec{x}')G(\vec{x}-\vec{x}')\,\d^{3}\vec{x}'.$
Substituting expression (\ref{eq:j-psi}) into the last integral,
after some algebra we obtain $\vec{a}=\mu_{0}\vec{\nabla}\times\chi\vec{e}_{z},$
where \[
\chi(\vec{x})=\int_{V}\psi(\vec{x}')G(\vec{x}-\vec{x}')\,\d^{3}\vec{x}'\]
is the same as used by Thess \emph{et al.}\cite{TVKZ-07} The previous
expression implies that $\vec{a},$ similarly to its source $\vec{j},$
has no $z$-component. Then the induced magnetic field can be written
as \[
\vec{b}=\mu_{0}\vec{\nabla}\times\vec{\nabla}\times\chi\vec{e}_{z}=\mu_{0}(\vec{e}_{z}\psi-\vec{\nabla}\phi),\]
where $\phi=-\partial_{z}\chi$ represents the scalar magnetic potential,
which completely defines $\vec{b}$ outside the layer, where $\psi=0.$
Further, using the tensor notation with the Einstein summation convention,
expressions (\ref{eq:F}) and (\ref{eq:M}) can be represented in
terms of $\chi$ as \begin{eqnarray}
F_{i} & = & \mu_{0}m_{j}\chi_{;ij3},\label{eq:Fi}\\
M_{i} & = & \mu_{0}\epsilon_{ijk}m_{j}\chi_{;k3},\label{eq:Mi}\end{eqnarray}
 where $\epsilon_{ijk}$ is the anti-symmetric tensor and the subscript
after the semicolon denotes the differential with respect to the corresponding
coordinate with indices $1,$ $2$ and $3$ standing for the $x,$$y$
and $z,$ directions, respectively. Taking into account the symmetry
of $G(\vec{x}-\vec{x}')$ with respect to the interchange of observation
and integration points, the derivatives of $\chi$ at the dipole location
$\vec{x}=0$ in expressions (\ref{eq:Fi}) and (\ref{eq:Mi}) are
found as \begin{eqnarray}
\chi_{;ij} & = & \int_{V}\psi(\vec{x})G_{;ij}(\vec{x})\,\d^{3}\vec{x},\label{eq:X-ij}\\
\chi_{;ijk} & = & -\int_{V}\psi(\vec{x})G_{;ijk}(\vec{x})\,\d^{3}\vec{x},\label{eq:X-ijk}\end{eqnarray}
 where \begin{eqnarray*}
G_{;ij}(\vec{x}) & = & \frac{3x_{i}x_{j}-\delta_{ij}|\vec{x}|^{2}}{4\pi|\vec{x}|^{5}},\\
G_{;ijk}(\vec{x}) & = & -\frac{15x_{i}x_{j}x_{k}-3(\delta_{ij}x_{k}+\delta_{ik}x_{j}+\delta_{jk}x_{i})|\vec{x}|^{2}}{4\pi|\vec{x}|^{7}}.\end{eqnarray*}
 Integrals (\ref{eq:X-ij}) and (\ref{eq:X-ijk}) can be evaluated
analytically using, for example, the computer algebra system Mathematica,\cite{Wolf96}
which also allows us to carry out all other analytical transformations.
In such a way, we firstly verify that Eq. (\ref{eq:Fi}) with expression
(\ref{eq:psi0-H}) substituted into integral (\ref{eq:X-ijk}) indeed
reproduces previous results (\ref{eq:Fx}), (\ref{eq:Fy}) and (\ref{eq:Fz}).
Secondly, Eq. (\ref{eq:Mi}) with the same expression for $\psi_{0}$
substituted in integral (\ref{eq:X-ij}) results in\begin{equation}
\vec{M}_{0}=\frac{\mu_{0}^{2}\sigma v}{128\pi h^{2}}\left(-m_{x}m_{y}\vec{e}_{x}+(m_{x}^{2}+m_{z}^{2})\vec{e}_{y}-m_{y}m_{z}\vec{e}_{z}\right),\label{eq:M0}\end{equation}
 which for a transversal dipole again coincides with the results of
Thess \emph{et al.}\cite{TVKZ-07}

\subsubsection{Rotation}

For a solid-body rotation with the velocity $\vec{v}_{1}=-\vec{\Omega}\times\vec{x},$
which appears in the co-rotating frame of reference, the charge conservation
$\vec{\nabla}\cdot\vec{j}=0$ applied to Eq. (\ref{eq:Ohm}) results
in\begin{equation}
\vec{\nabla}^{2}\varphi_{1}=-2\Omega B_{y},\label{eq:phi1}\end{equation}
 while the vanishing of the normal current component at the surface
$z=h$ requires\begin{equation}
\partial_{z}\varphi_{1}=-\Omega zB_{y}.\label{bc:phi1}\end{equation}
 In order to solve this problem, firstly, it is important to notice
that since $\int B_{y}\,\d z$ satisfies the Laplace equation, \begin{eqnarray}
\bar{\varphi}_{1} & = & \Omega z\int B_{y}\,\d z=\mu_{0}m\Omega z\partial_{ym}^{2}H\label{eq:phi1-b}\end{eqnarray}
 is a particular solution to Eq. (\ref{eq:phi1}). Comparing expressions
(\ref{eq:phi1-b}) and (\ref{sol:phi0}) shows that $\bar{\varphi}_{1}=(\Omega z/v)\varphi_{0},$
where $\varphi_{0}$ is given by solution (\ref{sol:phi0-H}). Searching
for the solution as $\varphi_{1}=\bar{\varphi}_{1}+\tilde{\varphi}_{1},$
reduces Eq. (\ref{eq:phi1}) to the Laplace equation for $\tilde{\varphi}_{1},$
while BC (\ref{bc:phi1}) takes the form \[
\partial_{z}\tilde{\varphi}_{1}=-\mu_{0}m\Omega\partial_{ym}H.\]
 By the usual arguments, we obtain \begin{equation}
\tilde{\varphi}_{1}=-\mu_{0}m\Omega\partial_{ym}H_{1},\label{eq:phi1-t}\end{equation}
 where $H_{1}=\int H\,\d z=\frac{1}{4\pi}\left(RQ_{1}(\cos\theta)+z\ln r\right)=zH-R^{2}G$
and $Q_{1}(z)=zQ_{0}(z)-1$ is the first order associated Legendre
function of the second kind.\cite{AbSt72} Using the expressions
above, after some transformations we obtain\begin{eqnarray}
\varphi_{1} & = & -\mu_{0}\Omega\left(\vec{e}_{x}\cdot\vec{m}\times\vec{\nabla}H+ym\partial_{m}G\right)=\nonumber \\
 &  & -\frac{\mu_{0}\Omega}{4\pi}\left[\frac{m_{z}(R^{3}-z^{3})\sin\vartheta}{R^{3}r}+\frac{(m_{x}\sin2\vartheta-m_{y}\cos2\vartheta)r^{2}-m_{y}(R^{2}+z^{2})}{2R^{3}}\right].\label{sol:phi1}\end{eqnarray}
 which is the electric potential in the co-rotating frame of reference.

\begin{figure}
\begin{centering}
\includegraphics[width=0.5\columnwidth]{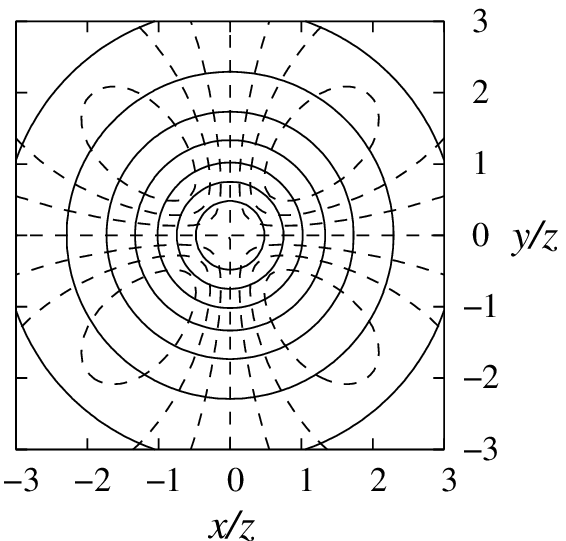}\put(-30,0){(a)}\includegraphics[width=0.5\columnwidth]{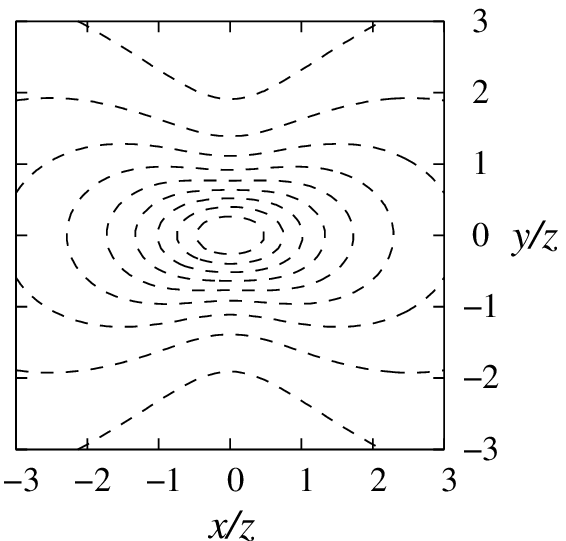}\put(-30,0){(b)}
\par\end{centering}

\begin{centering}
\includegraphics[width=0.5\columnwidth]{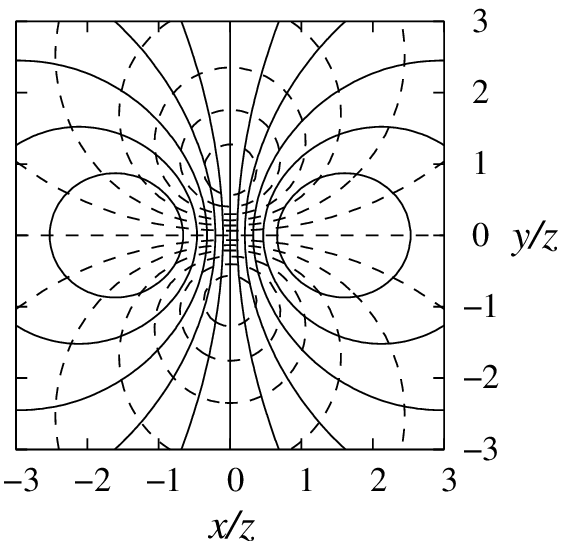}\put(-30,0){(c)} 
\par\end{centering}

\caption{\label{fig:rot-xyz}Instantaneous isolines of the electric potential
$\varphi_{1}$ (dashed) and of the stream function $\psi_{1}$ (solid),
which represent the eddy current lines, induced by a solid-body rotation
in the magnetic field of the dipole aligned with the $x$- (a), $y$-
(b), and $z$-axis (c).}

\end{figure}

As before, condition (\ref{bc:phi1}) is satisfied not only at $z=h$
but any $z$ and, thus, the $z$-component of the current vanishes
throughout the layer. Consequently, the electric current can again
be expressed as (\ref{eq:j-psi}) with $\psi_{1},$ for which Eq.
(\ref{eq:psi-gen}) takes the form\begin{equation}
(\vec{\nabla}^{2}-\partial_{z}^{2})\psi_{1}=\sigma\Omega(B_{x}+\vec{e}_{y}\cdot\vec{r}\times\vec{\nabla}B_{z}).\label{eq:psi1}\end{equation}
 Since the r.h.s. of Eq. (\ref{eq:psi1}) satisfies the Laplace equation,
after some algebra we obtain\begin{eqnarray}
\psi_{1} & = & -\sigma\Omega\left(\iint B_{x}\,\d z^{2}+\vec{e}_{y}\cdot\vec{r}\times\vec{\nabla}\iint B_{z}\,\d z^{2}-2\partial_{x}\iiint B_{z}\,\d z^{3}\right)\nonumber \\
 & = & \mu_{0}\sigma\Omega\vec{e}_{y}\cdot\vec{m}\times\vec{\nabla}H=\frac{\mu_{0}\sigma\Omega}{4\pi}\left(\frac{m_{z}(R-z)\cos\vartheta-m_{x}r}{Rr}\right).\label{sol:psi1}\end{eqnarray}
 The eddy current lines in the $xy$-plane, which are the isolines
of $\psi_{1},$ along with the isolines of the electric potential
(\ref{sol:phi1}) are shown in Fig. \ref{fig:rot-xyz}. Again, the
patterns are self-similar and scale directly with the distance $z$
from the dipole, while the magnitude of both $\psi_{1}$ and $\varphi_{1}$
falls off as $\sim z^{-1}.$ In contrast to the translation considered
above, the eddy currents induced by rotation are time-dependent and
vary periodically as the dipole orientation changes from the $x$-
to the $z$-axis, which are shown in Figs. \ref{fig:rot-xyz}(a) and
(c), respectively. Only the electric potential, but no current, is
induced by a solid-body rotation when the dipole is aligned with the
axis of rotation ($y$). In his case, when the magnetic field is symmetric
about the axis of rotation, the induced e.m.f. is irrotational and,
thus, compensated by the electric potential gradient, which equals
to \begin{equation}
-\vec{v}_{1}\times\vec{B}=-\Omega\rho\vec{\epsilon}\times\vec{\nabla}\times(A\vec{\epsilon})=-\Omega\vec{\nabla}(\rho A),\label{eq:v1xB}\end{equation}
 where $\rho=|\vec{\Omega}\times\vec{x}|/\Omega$ is the cylindrical
radius from the axis of rotation, $\vec{\epsilon}=\vec{\Omega}\times\vec{x}/|\vec{\Omega}\times\vec{x}|$
is the azimuthal unity vector, and $A$ is the azimuthal component
of the vector potential, which can be used to describe a general axially
symmetric poloidal magnetic field. For the dipole field, we have\[
\vec{A}(\vec{x})=\mu_{0}\int_{\bar{V}}\vec{J}(\vec{x}')G(\vec{x}-\vec{x}')\,\d^{3}\vec{x}'=\vec{m}\times\vec{\nabla}G(\vec{x})=A(\vec{x})\vec{\epsilon},\]
 where $\vec{J}$ is the associated dipole current (\ref{eq:J}) and
$A(\vec{x})=-|\vec{m}\times\vec{x}|G(\vec{x})/|\vec{x}|^{2}.$ Substituting
this into expression (\ref{eq:v1xB}) and equating it to the gradient
of the electric potential $\varphi_{1},$ we obtain \[
\varphi_{1}=-\Omega\rho A=\frac{\mu_{0}m_{y}\Omega(R^{2}-y^{2})}{4\pi R^{3}},\]
 which is the free-space electric potential in the co-rotating frame
of reference shown in Fig. \ref{fig:rot-xyz}(b) and coinciding with
(\ref{sol:phi1}) when $m_{x}=m_{z}=0.$ This potential vanishes in
the laboratory frame of reference, where the magnetic field is invariant
with respect to the rotation around the symmetry axis.

Further, using expression (\ref{sol:psi1}) for $\psi_{1}$ in integrals
(\ref{eq:X-ij}) and (\ref{eq:X-ijk}), which can be evaluated together
with Eqs. (\ref{eq:Fi}) and (\ref{eq:Mi}) in the same way as for
the translation in the previous section, we obtain \begin{eqnarray}
\vec{F}_{1} & = & \frac{\mu_{0}^{2}\sigma\Omega}{128\pi h^{2}}(-(m_{x}^{2}+m_{z}^{2})\vec{e}_{x}-m_{x}m_{y}\vec{e}_{y}+m_{x}m_{z}\vec{e}_{z}),\label{eq:F1}\\
\vec{M}_{1} & = & \frac{\mu_{0}^{2}\sigma\Omega}{64\pi h}(2m_{x}m_{y}\vec{e}_{x}-(2m_{x}^{2}+m_{z}^{2})\vec{e}_{y}+m_{y}m_{z}\vec{e}_{z}).\label{eq:M1}\end{eqnarray}

The first point to note is that both the force and torque vanishes
when dipole is aligned with the axis of rotation $(m_{x}=m_{z}=0),$
which, as discussed above, is due to the absence of eddy currents
in this case. Secondly, the longitudinal $(x)$ component of force
(\ref{eq:F1}), in contrast to that due to the translation given by
expression (\ref{eq:Fx}), is independent of the dipole orientation
in the $xz$-plane and depends only on the magnitude of the dipole
moment in this plane: $\bar{m}^{2}=m_{x}^{2}+m_{z}^{2}.$ Thirdly,
there is also a non-zero transversal $(z)$ force, which in contrast
to the longitudinal one is purely oscillatory and varies periodically
with the dipole orientation in the $xz$-plane as $m_{x}m_{z}=\frac{1}{2}\bar{m}^{2}\sin2\vartheta_{0},$
where $\vartheta_{0}$ is the poloidal angle of the dipole orientation
in the $xz$-plane from the positive $x$-axis, which is the same
as that for the linear dipole in expression (\ref{eq:A2d}) (see Fig.
\ref{cap:Sketch}). Moreover, when dipole is inclined to the axis
of rotation ($m_{x}m_{y}\not=0),$ also a spanwise $(y)$ force appears,
which similarly to the transversal one alternates periodically around
zero as $m_{x}m_{y}=\bar{m}m_{y}\cos\vartheta_{0}$ with the dipole
rotation. As seen from expression (\ref{eq:M1}), inclination also
gives rise to alternating torque components around both the $x$-
and $z$-axis. The $y$-component of torque, which is negative and,
thus, opposing the rotation, has not only a constant but also an alternating
part, which varies with the double frequency of rotation as $-(2m_{x}^{2}+m_{z}^{2})=-\frac{3}{2}\bar{m}^{2}\left(1-\frac{1}{3}\cos2\vartheta_{0}\right).$

\begin{figure}
\begin{centering}
\includegraphics[width=0.5\textwidth]{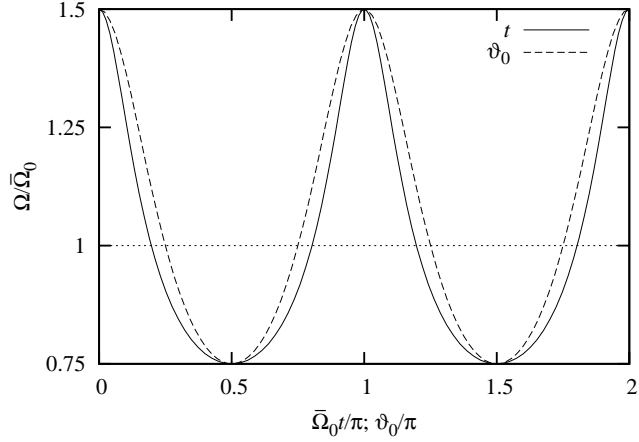} 
\par\end{centering}

\caption{\label{fig:omg-t}Instantaneous rotation rate (\ref{eq:omgt-trq})
versus time (solid) and orientation (dashed) for a negligible inertia
of the magnet.}

\end{figure}

Thus, in contrast to the long magnet in $\S$\ref{sub:2D}, there
is no steady rotation rate, which could balance the $y$-component
of the constant torque ($\ref{eq:M0}$) due to the translation. The
variation of the magnet rotation rate in response to oscillatory torque
is constrained by the inertia of magnet. For the oscillatory part
of the rotation rate $\tilde{\Omega}$ to be negligible compared to
the mean one $\bar{\Omega},$ we need \begin{equation}
\frac{\tilde{\Omega}}{\bar{\Omega}}\sim\frac{M}{\bar{\Omega}^{2}I}\sim\frac{\mu_{0}^{2}m^{2}\sigma}{\bar{\Omega}Ih}\ll1\label{eq:omeg-osc}\end{equation}
where $I$ is the moment of inertia of magnet. If this condition is
satisfied, the oscillatory component may be neglected in the balance
of torques (\ref{eq:M0}) and (\ref{eq:M1}) around the axis of rotation,
which is the $y$-axis. For a layer of finite thickness, this results
in \begin{equation}
\bar{\Omega}_{0}=\frac{v}{3}\left(\frac{1}{h_{1}}+\frac{1}{h_{2}}\right).\label{eq:omgb-trq}\end{equation}
 The expression above differs only by a factor of $\frac{2}{3}$ from
result (\ref{eq:omg-frc}) for a long magnet in the case of a vanishing
longitudinal force. In the opposite limit of a negligible inertia,
the torque balance yields\begin{equation}
\Omega=\bar{\Omega}_{0}/(1-\tfrac{1}{3}\cos2\vartheta_{0}).\label{eq:omgt-trq}\end{equation}
This instantaneous rotation rate $\Omega=\partial_{t}\vartheta_{0},$
which is shown in Fig. \ref{fig:omg-t} versus both the time $t$
and orientation $\vartheta_{0},$ alternates between $\frac{3}{4}\bar{\Omega}_{0}$
and $\frac{3}{2}\bar{\Omega}_{0}$ of the mean value (\ref{eq:omgb-trq})
$\bar{\Omega}_{0}=\left.\vartheta_{0}/t\right|_{t\rightarrow\infty}.$
The last relation follows from the integration of Eq. (\ref{eq:omgt-trq})
as $\vartheta_{0}-\frac{1}{6}\sin2\vartheta_{0}=\bar{\Omega}_{0}t,$
which also defines $\Omega$ parametrically versus $t$ in Fig. \ref{fig:omg-t}.
Note that the oscillations of the rotation rate do not affect its
mean value $\bar{\Omega}_{0},$ which is defined by (\ref{eq:omgb-trq})
independently of condition (\ref{eq:omeg-osc}). 

Alternatively, when the magnet rotation rate is actively controlled
so that to balance the longitudinal force (\ref{eq:Fx}) with the
$x$-component of force (\ref{eq:F1}), the instantaneous rotation
rate is \begin{equation}
\Omega=\partial_{t}\vartheta_{0}=\frac{\bar{\Omega}_{1}}{4\sqrt{3}}(7-\cos2\vartheta_{0}),\label{eq:omgt-frc}\end{equation}
where the temporal mean value \begin{equation}
\bar{\Omega}_{1}=\frac{\sqrt{3}v}{2}\left(\frac{1}{h_{1}}+\frac{1}{h_{2}}-\frac{1}{h_{1}+h_{2}}\right)\label{eq:omgb-frc}\end{equation}
follows from the solution of differential equation (\ref{eq:omgt-frc}),
which can be written as \[
\tan(\vartheta_{0}(t))=\frac{\sqrt{3}}{2}\tan(\bar{\Omega}_{1}t).\]

\section{\label{sec:Exp}Implementation and testing}

\begin{figure}
\begin{centering}
\includegraphics[width=0.5\textwidth]{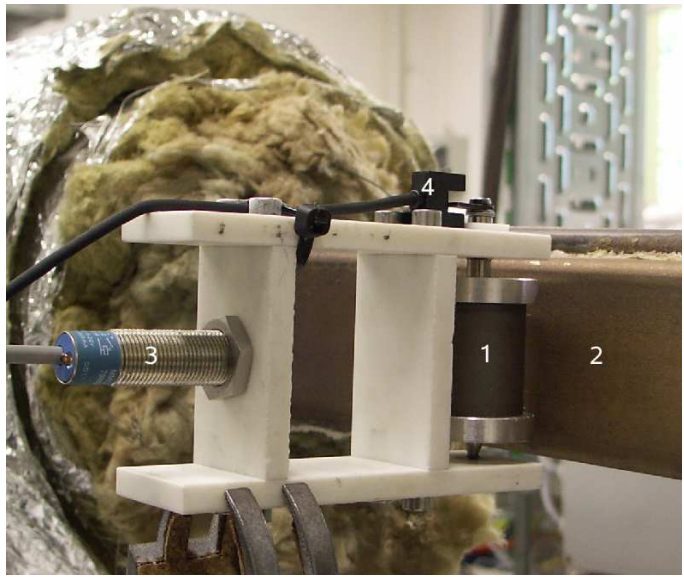} 
\par\end{centering}

\caption{\label{exp-pic}Single-magnet rotary flowmeter with a \emph{SmCo}-type
magnet (1) of $24\,\mbox{mm}$ and $35\,\mbox{mm}$ in diameter and
height, respectively, installed at the side of a rectangular stainless
steel duct (2) of $45\times45\,\mbox{mm}^{2}$ in cross-section carrying
the liquid sodium flow. The rotation rate is measured with an inductive
magnetic proximity sensor (3) and also monitored with a photoelectric
barrier sensor (4). }

\end{figure}

The laboratory model of the single-magnet flowmeter shown in Fig.
\ref{exp-pic} consists of a cylindrical \emph{SmCo}-type permanent
magnet with diameter $2R=24\,\mbox{mm}$ and height $L=35\,\mbox{mm}$,
magnetized perpendicularly to its axis with $0.5\,\mbox{T}$ surface
induction, which holds up to $230^{\circ}\mbox{C}.$ The magnet is
mounted on a stainless steel axle held by a housing made of {}``MACOR''
ceramic, which ensures a low mechanical friction and can withstand
up to $800^{\circ}\mbox{C}.$

The flowmeter was tested on a sodium loop using about 90 liters of
molten \emph{Na} at $220^{\circ}\mbox{C}$ with the electrical conductivity
of $9\times10^{6}\,\mbox{S}\,\mbox{m}^{-1}.$ The flow was driven
by a linear electromagnetic pump, which provided a maximum velocity
of $1.5\,\mbox{m}\,\mbox{s}^{-1}$ at the flow rate of $3\,\mbox{l}\,\mbox{s}^{-1}$
in a stainless steel duct of $45\times45\,\mbox{mm}^{2}$ cross-section.

The frequency of magnet rotation was determined with an inductive
magnetic proximity sensor (SICK MM12-60APS-ZU0), which was fixed at
the distance of $50\,\mbox{mm}$ perpendicularly to the side surface
of magnet at its midheight as shown in Fig. \ref{exp-pic}. The sensor
briefly switched off and then on again as the magnetic field along
the axis of the sensor changed its direction. The frequency of sensor
transition from the off to the on state, which happens twice per revolution,
was measured with Keithley 2000 Multimeter using the reciprocal frequency
counting techniques with the measurement gate time set to $1\,\mbox{s}.$
The frequency was also monitored with Tektronix TDS 210 oscilloscope.
After changing the pump current, the flow was allowed to develop for
$2\,\mbox{min},$ and then six measurements were taken with the intervals
of approximately $30\,\mbox{s}.$ The standard deviation in the measured
rotation frequencies was typically a few tenths of $\mbox{Hz.}$

\begin{figure}
\begin{centering}
\includegraphics[width=0.5\textwidth]{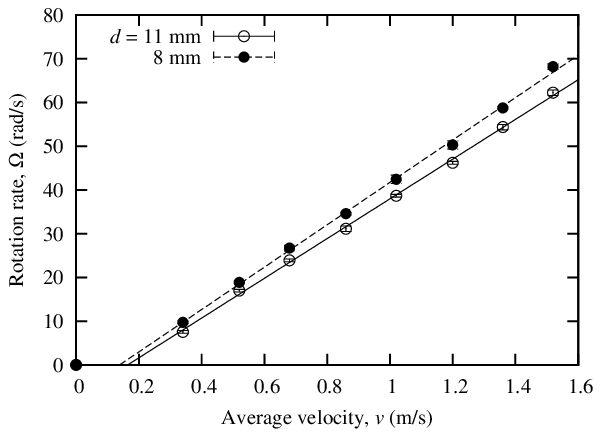}\put(-30,30){(a)}\includegraphics[width=0.5\textwidth]{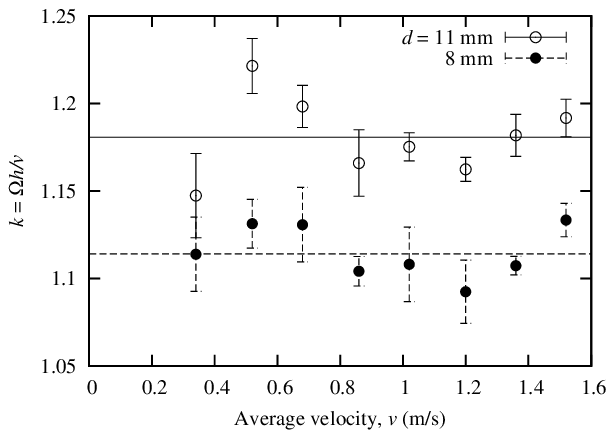}\put(-30,30){(b)} 
\par\end{centering}

\caption{\label{fig:exp}Measured (a) and rescaled (b) magnet rotation rates
along with the best linear fits versus the average velocity for $d=8\,\mbox{mm}$
and $11\,\mbox{mm}$ gaps between the magnet and the duct with the
distance between the magnet axis and liquid metal $h=R+d_{w}+d,$
where $R=12\,\mbox{mm}$ is the magnet radius and $d_{w}=3\,\mbox{mm}$
is the thickness of stainless steel wall.}

\end{figure}

The rotation rates measured at two gap widths between the magnet and
the duct are shown in Fig. \ref{fig:exp}(a) depending on the average
sodium velocity, which was determined using Ultrasonic Doppler Velocimetry\cite{Eck-etal-11}
with the resolution of $1\,\mbox{cm}\,\mbox{s}^{-1}.$\cite{EG-02}
Although the rotation rate is seen to increase as the magnet is approached
to the duct and to vary nearly linearly with the flow velocity in
agreement with the theory, the linear fit shows a certain zero offset.
Obviously, there is some opposing force, which has to be overcome
for the rotation to start. This requires the flow velocity of at least
$0.2\,\mbox{m}\,\mbox{s}^{-1}.$ First, such an opposing force is
caused by the dry friction in bearings. Second, the magnet is oriented
by the Earth's magnetic field, which has a stronger effect than the
friction and also has to be overcome for the rotation to start. Third,
even stronger effect is caused by the stainless steel walls of the
duct, which are weakly ferromagnetic.\cite{WebFaj-98} The magnet,
when approached to the duct, was observed to turn with the dipole
moment perpendicular to the wall. It is also important to note that
the magnet, when turned by hand away from the equilibrium orientation,
returned to it monotonically without oscillations. This implies the
inertia of magnet to be small relative to the electromagnetic drag
force. Thus, the inertia cannot significantly contribute to the overcoming
of the orienting torque due to the wall magnetization. However, when
magnet rotates slowly, the orienting torque causes the rotation rate
to fluctuate, which shows up as an increased scatter in the measured
values seen in Fig. \ref{fig:exp} at low flow velocities. 

When the zero offset is removed and the remaining rotation rate is
multiplied by the distance $h=R+d_{w}+d$ between the magnet axis
and the liquid metal, where $R=12\,\mbox{mm}$ is the magnet radius
and $d_{w}=3\,\mbox{mm}$ is the thickness of stainless steel wall,
we obtain the relative rotation rate $k=\Omega h/v,$ which represents
a rescaled slope coefficient for Fig. \ref{fig:exp}(a) and is plotted
in Fig. \ref{fig:exp}(b) versus the velocity. The short-magnet solution
(\ref{eq:omgb-trq}) in two limiting cases of a semi-infinite $(h_{2}=\infty)$
and a thin ($h_{2}=h_{1})$ layer yields $k=\frac{1}{3}$ and $k=\frac{2}{3},$
respectively, which are smaller than the measured values $k\approx1$
seen in Fig. \ref{fig:exp}(b). On the other hand, the long-magnet
solution (\ref{eq:omg-trq}), which appears more adequate for the
experimental setup, yields $k\le\frac{1}{2}$ for $h_{1}\le h_{2}.$
Such quantitative differences between the experiment and theory are
not surprising given the simplifications underlying the latter. First,
theoretical model does not take into account the finite width of the
duct. But this alone could hardly explain the high rotation rate of
the magnet at which its field travels faster than the layer. This
apparently being the case in the experiment implies the presence of
significant velocity gradients, which are also ignored by the theory,
but could be taken into account as outlined in Sec. \ref{sub:2D}.
Note that strong velocity gradients can be caused by the magnet itself,
which due to its large size and strength may act as a magnetic obstacle
partially blocking the flow and so increasing its local velocity.

\section{\label{sec:Con}Summary and conclusions}

We have presented a theory of single-magnet rotary flowmeter for two
limiting cases of long and short magnets, which were modeled as linear
and single dipoles, respectively. Simple analytical solutions were
obtained for the force and torque on slowly translating and rotating
magnets due to eddy currents in layers of arbitrary depth and infinite
horizontal extent. The velocity was assumed to be constant and the
motion so slow that the induced magnetic field could be neglected.
The latter assumption was not applicable to the long magnet rotating
above a conducting half-space. In this case, to obtain a finite braking
torque, the skin effect due to the induced magnetic field had to be
taken into account however slow the rotation. For a single dipole
of arbitrary orientation, compact analytical solutions were obtained
in terms of both the electric potential and stream function induced
by the layer translation and rotation in the co-rotating frame of
reference. The electric stream function was used further to find the
scalar potential of induced magnetic field at the dipole location,
which resulted in simple expressions for the force and torque on the
dipole. Eventually, we found the equilibrium rotation rate at which
the driving torque due to the layer translation is balanced by the
braking torque due to the magnet rotation. An alternative approach
was also considered, where the force on the magnet could be used to
control its rotation rate so that the resulting force vanishes. In
either case, the resulting rotation rate is directly proportional
to the layer velocity and inversely proportional to the distance between
the magnet and the liquid metal. These results were found in a qualitative
agreement with the measurements on the liquid sodium flow. A more
accurate quantitative agreement with experiment is limited due to
the substantial approximations underlying the theoretical model, which
neglects the finite lateral extension of the layer as well as the
spatial and temporal variations of the velocity distribution.

In conclusion, note that the resulting rotation rate is independent
of the magnet strength and the electrical conductivity of the liquid
metal provided that the mechanical friction or other external effects
are negligible compared to the driving torque. This is the main advantage
of rotary flowmeter over the LFV.\cite{TVK06}

\begin{acknowledgments}
J.P. has benefited from a stimulating discussion with Y. Kolesnikov
and would like to thank X. Wang for bringing Ref. {[}13] to his attention.
Experiments at FZDR were supported by a BMWi/AiF project. 
\end{acknowledgments}

\end{document}